\documentclass[twocolumn,jcp,aip,amssymb,amsmath,reprint]{revtex4-1}
\usepackage{graphicx}
\usepackage{epstopdf}
\usepackage{bm}
\usepackage{hyperref}
\usepackage{cleveref}
\usepackage{comment}
\usepackage{enumitem}

\hypersetup{%
   pdfpagemode=None, 
   pdfstartpage=1,
   pdfmenubar=true,
   pdftoolbar=true,
   colorlinks = true,
   linkcolor=blue,
   citecolor=blue,
   urlcolor=blue,
   bookmarksopen=false
 }
 
\crefname{figure}{fig.}{figs.}
\Crefname{figure}{Figure}{Figures}
\Crefname{section}{Section}{Sections}
\Crefname{equation}{Eq.}{Eqs.}

\begin{document}

\title{Long-range interactions of aromatic molecules with alkali-metal and alkaline-earth-metal atoms}  

\author{Leonid Shirkov}
\affiliation{Institute of Physics, Polish Academy of Sciences, Al.~Lotnik\'ow 32/46, 02-668 Warsaw, Poland}
\author{Micha{\l} Tomza}
\email{michal.tomza@fuw.edu.pl}
\affiliation{Faculty of Physics, University of Warsaw, Pasteura 5, 02-093 Warsaw, Poland}

\begin{abstract}
The isotropic and anisotropic coefficients $C^{l,m}_n$  of the long-range spherical expansion $\sim 1/R^n$ ($R$ -- the intermolecular distance) of the dispersion and inductions intermolecular energies are calculated using the first principles for the complexes containing an aromatic molecule
(benzene, pyridine, furan, and pyrrole) and alkali-metal (Li, Na, K, Rb, and Cs) or alkaline-earth-metal (Be, Mg, Ca, Sr, and Ba) atoms in their electronic ground states. The values of the first- and second-order properties of the aromatic molecules are calculated using the response theory with the asymptotically corrected LPBE0 functional. The second-order properties of the closed-shell alkaline-earth-metal atoms are obtained using the expectation-value coupled cluster theory and of the open-shell alkali-metal atoms using analytical wavefunctions. These properties are used for the calculation of the dispersion $C^{l,m}_{n,\text{disp}}$ and induction $C^{l,m}_{n,\text{ind}}$ coefficients ($C^{l,m}_n=C^{l,m}_{n,\text{disp}}+C^{l,m}_{n,\text{ind}}$) with $n$ up to 12 using the available implemented analytical formulas.  It is shown that the inclusion of the coefficients with $n>6$ is important for reproducing the interaction energy in the van der Waals region at $R\approx 6$ \AA. The reported long-range potentials should be useful for constructing the analytical potentials, valid for the whole intermolecular interaction range, which are needed for spectroscopic and scattering studies.      
\end{abstract}

\maketitle

\pagestyle{plain}

\section{Introduction\label{intr}}

The study of cold ($<$1 K) and ultracold ($<$1 mK) molecules open prospects in many areas of modern chemical physics and physical chemistry. For example, ultracold molecules provide a convenient platform for high-resolution spectroscopy, \cite{NesbittCR12} external control of chemical reactions, \cite{QuemenerCR12} and tests of fundamental theories.\cite{SafronovaRMP18} In order to cool molecules to the ultracold regime, they first have to be trapped in electric and magnetic fields by one of numerous existing techniques.\cite{Carr:2009,Prehn:2016,Kozyryev:2017_1} Next, sympathetic cooling\cite{Wallis:2009,Morita:2017,TomzaPCCP17,SonNature19} can potentially be used in the second stage to achieve even lower sub-mK temperatures by collisions with ultracold atoms, in which the elastic-to-inelastic rate ratio should be greater than 100.\cite{Carr:2009}

Recently there has been an increasing experimental interest in the cooling of polyatomic molecules.\cite{Kozyryev:2016,Mitra:2020} Possession of additional rotational and vibrational degrees of freedom by polyatomic molecules opens new prospects for many applications ranging from quantum computation\cite{AlbertPRX20} and simulation\cite{Wall:2015} to high-precision measurements\cite{Kozyryev:2017_2} and ultracold organic chemistry.\cite{Ivanov:2020} On the other hand, the additional degrees of freedom bring up new challenges both for experiment and theory. The density of vibrational energy levels imposes a desirable requirement on the diagonality of Franck-Condon factors,  though at the same time, complicates cooling due to energy loss resulting from inelastic collisions. The theoretical description of such processes requires careful treatment or the development of new theoretical methods.\cite{Ivanov:2020}

A big advantage of noble-gas (Rg) atoms as the collision partner is their inertness. The scheme of using such a technique to cool molecules was proposed theoretically some time ago.\cite{Barletta:2008, Barletta:2009,Barletta:2010} Even though such a scheme seems promising, only cryogenic temperatures could be achieved. Therefore, second-stage sympathetic cooling by collisions with polarizable species such as ultracold alkali-metal and alkaline-earth-metal atoms remains currently the important research goal. The theoretical study of such complexes is more difficult than that of the complexes with Rg atoms due to the higher anisotropy of the potentials, larger binding energies, and more distinct long-range behavior of the dispersion energy. Nevertheless, it was recently experimentally shown that the optical cycling of functionalized polyatomic molecules containing aromatic rings is feasible.\cite{AugenbraunJPCL22,MitraJPCL22} This opens the way to laser cooling and trapping such molecules at cold conditions,\cite{MitraScience20,VilasNature22} and next to sympathetic cooling them with ultracold metal atoms to even lower temperatures and application, e.g., in ultracold organic chemistry.\cite{Ivanov:2020} Such studies will require detailed knowledge of intermolecular interactions, including long-range potentials.

Besides the prospects for sympathetic cooling of aromatic molecules, there are other reasons for the interest in such complexes. The complexes of pyridine with Li, Ca, and Sr have already been studied by means of laser spectroscopy. \cite{Krasnokutski:2009} The authors used the zero-electron-kinetic-energy spectroscopy to determine binding and ionization energies as well as intermolecular vibrational energies.  The assignment of the energies requires an accurate potential that will be used for the solution of the three-dimensional rovibrational Schr{\"o}dinger equation.\cite{Shirkov:2020}

The intermolecular interaction energy  $E_{\text{int}}$  can be effectively separated into terms describing different physical effects such as electrostatic, induction, dispersion, and exchange. The most rigorous approach to studying such a separation is based on the symmetry-adapted perturbation theory (SAPT). \cite{Jeziorski:1994} When the distance $R$ between the monomer centers of mass is large, the exchange effect vanishes, and the potential can effectively be described by the spherical multipole expansion, which contains terms proportional to $1/R^n$,\cite{Avoird:1980,Rijks:1989_1,Rijks:1989_2}including the first-order electrostatic energy $E_{\text{elst}}^{(1)}$ and the second-order dispersion $E_{\text{disp}}^{(2)}$ and induction $E_{\text{ind}}^{(2)}$ energies. These series become divergent for smaller $R$ values and the interaction energy  $E_{\text{int}}$ defined only through a multipole expansion tends to unphysically negative values. However,   such a long-range potential is expected to be applicable for the values of $R$ close to the equilibrium if the number of expansion terms is large enough. For example, the inclusion of the long-range dispersion coefficients $C_n$ up to $n\!=\!12$ for Rg dimers accurately reproduces the second-order dispersion  $E_{\text{disp}}^{(2)}$ even at the equilibrium configurations.\cite{Shirkov:2017}  Moreover, the long-range series can be useful for the approximate methods 
in which the non-dispersion terms are found  either at Hartree-Fock or dispersionless DFT level.\cite{Pernal:2009_2,Podeszwa:2010_3}  In this case the dispersion terms  are multiplied by the appropriate damping functions in the short-range region.\cite{Knowles:1986,Misquitta:2005_2,Moszynski:2007} This approach gave a rise to SAPT+D methods \cite{Hesselmann:2011_1, Lao:2013, Jacobson:2013} an a widely used dispersion-corrected density functional theory (DFT-D).\cite{Grimme:2011,Caldeweyher:2019} 

The complexes of aromatic molecules with polarizable atoms, such as alkali-metal and alkaline-earth-metal atoms, when compared to the complexes of aromatic molecules with Rg atoms,\cite{Shirkov:2015_2,Shirkov:2020} are characterized by higher magnitudes of the polarization terms due to larger values of polarizabilities of the metal atoms. Therefore, an accurate description of the long-range potentials for such complexes is crucial for creating reliable potentials valid in the whole physically relevant interaction range. An initial theoretical study of the complexes of aromatic molecules with alkali-metal and alkaline-earth-metal atoms has already been performed by one of us. \cite{Wojcik:2019} That study focused on the complexes with aromatic molecules such as benzene, naphthalene, and azulene. It included the calculation of the leading long-range dispersion and induction coefficients as well as the construction of analytical potentials derived from {\slshape ab initio} CCSD(T) calculations. Other theoretical works on similar complexes mostly focused on studying equilibrium structures.\cite{Denis:2013,Denis:2014,Sadlej:2015,Borca:2016,Ullah:2018}

The goal of the present work is to improve the long-range potential for the complexes with benzene by including more terms in the multipole expansion series than had been done using the first principles methods. \cite{Wojcik:2019} The study of the complexes with benzene is important because benzene is a prototypical molecule of high symmetry. However, the lack of the electric dipole moment complicates its experimental study in external fields. Therefore, a few other typical single-ring aromatic molecules with non-zero dipole moments, such as pyridine, furan, and pyrrole, are also investigated in our study. In this way, our work establishes the computational scheme for calculating the long-range interaction coefficients between small aromatic molecules and polarizable metal atoms, which can be useful for constructing potential energy surfaces (PESs) between laser-coolable functionalized aromatic molecules\cite{AugenbraunJPCL22,MitraJPCL22} and metal atoms for scattering calculations to predict and guide sympathetic cooling in upcoming experiments at ultralow temperatures.\cite{MitraScience20,VilasNature22}

This article is organized as follows. In \Cref{comp}  the theoretical and computational methods employed in this work are explained. \Cref{res} presents the results,  and \Cref{conc} contains the conclusions of the work and future prospects.
\\

\section{Computational and Theoretical Methods\label{comp}}

\subsection{Long-range potentials\label{long}}

Let us consider two interacting non-linear molecules A and B in their electronic ground states.
The interaction energy in the long range is defined by the multipole expansion:\cite{Avoird:1980,Rijks:1989_2}
\begin{equation}
\begin{split}
E_{\text{int}}(\boldsymbol{R},\boldsymbol{\omega}_{\text{A}},\boldsymbol{\omega}_{\text{B}})=\sum_{L_{\text{A}} K_{\text{A}} L_{\text{B}} L_{\text{B}}L}   \frac{C_n^{L_{\text{A}} K_{\text{A}} L_{\text{B}} L_{\text{B}}L}}{R^n} \times\\
\times A_{L_{\text{A}} K_{\text{A}} L_{\text{B}} L_{\text{B}}L}(\boldsymbol{\omega}_{\text{A}},\boldsymbol{\omega}_{\text{B}},\boldsymbol{\Omega})\,,
\label{eq:eint}
\end{split}
\end{equation}
where $\boldsymbol{R}=(R,\boldsymbol{\Omega})=(R,\Theta,\Phi)$ is the vector connecting the monomer centers of mass, and the orientations of the molecules A and B are described by the Euler angles $\boldsymbol{\omega}_{\text{A}}=(\alpha_{\text{A}},\beta_{\text{A}},\gamma_{\text{A}})$ and $\boldsymbol{\omega}_{\text{B}}=(\alpha_{\text{B}},\beta_{\text{B}},\gamma_{\text{B}})$ in the molecule-fixed reference frame. The angular functions are given by:
\begin{equation}
\begin{split}
A_{L_{\text{A}} K_{\text{A}} L_{\text{B}} L_{\text{B}}L}(\boldsymbol{\omega}_{\text{A}},\boldsymbol{\omega}_{\text{B}},\boldsymbol{\Omega})
=\sum_{M_{\text{A}}M_{\text{B}}M}
\begin{pmatrix}
L_{\text{A}}      & L_{\text{B}} & L \\
M_{\text{A}}      & M_{\text{A}} & M\\
\end{pmatrix}
\\
\times D^{L_{\text{A}}}_{M_{\text{A}}K_{\text{A}}}(\boldsymbol{\omega}_{\text{A}})^{\star} 
D^{L_{\text{B}}}_{M_{\text{B}}K_{\text{B}}}(\boldsymbol{\omega}_{\text{B}})^{\star}
C^{L}_{M}(\boldsymbol{\Omega}),
\label{eq:ang}
\end{split}
\end{equation}
where $D^{L}_{MK}(\boldsymbol{\omega})$ are the standard rotation Wigner matrices, $C^{L}_{M}(\boldsymbol{\Omega})$ are the Racah normalized spherical harmonics, and 
$\begin{pmatrix} 
L_{\text{A}}      & L_{\text{B}} & L \\
M_{\text{A}}      & M_{\text{A}} & M\\
\end{pmatrix}
$
are the 3-$j$ symbols.\cite{Avoird:1980,Zare:1988,Stone:2013}

The long-range general formulas for the calculation of $C_{n}^{L_{\text{A}},K_{\text{A}},L_{\text{B}},K_{\text{B}},L}$ had been derived before. \cite{Wormer:1977,Avoird:1980,Rijks:1988,Rijks:1989_2,Wormer:1992}  The electrostatic, induction, and dispersion coefficients are given by the following expressions: 
\begin{widetext}
\begin{equation}\label{eq:elst}
C_{n,\text{elst}}^{L_{\text{A}},K_{\text{A}},L_{\text{B}},K_{\text{B}},L}=(-1)^{L_{\text{A}}}\delta_{L_{\text{A}}+L_{\text{B}},L}\delta_{n,L+1}
\left[\frac{(2L_{\text{A}}+2L_{\text{B}}+1)!}{(2L_{\text{A}})!(2L_{\text{B}})!}\right]^{1/2}Q^{L_{\text{A}}}_{K_{\text{A}}}Q^{L_{\text{B}}}_{K_{\text{B}}},
\end{equation}
\begin{equation}\label{eq:ind}
\begin{split}
C_{n,\text{ind}}^{L_{\text{A}},K_{\text{A}},L_{\text{B}},K_{\text{B}},L}(\text{A} \rightarrow \text{B})=-\frac12
\sum_{l_{\text{A}} l_{\text{A}}^{\prime}l_{\text{B}} l_{\text{B}}^{\prime}}^{{l_{\text{A}}+l_{\text{A}}^{\prime}+l_{\text{B}}+l_{\text{B}}^{\prime}}+2=n}
\zeta^{L_{\text{A}}L_{\text{B}}L}_{l_{\text{A}} l_{\text{A}}^{\prime}l_{\text{B}} l_{\text{B}}^{\prime}}
\sum_{m_{\text{A}}=-l_{\text{A}}}^{l_{\text{A}}}(-1)^{K_{\text{A}}}
\begin{pmatrix}
l_{\text{A}}      & l_{\text{A}}^{\prime} & L_{\text{A}} \\
m_{\text{A}}      & K_{\text{A}}-m_{\text{A}} & -K_{\text{A}} \\
\end{pmatrix}\times\\
\times \sum_{m_{\text{B}}=-l_{\text{B}}}^{l_{\text{B}}}(-1)^{K_{\text{B}}}
\begin{pmatrix}
l_{\text{B}}      & l_{\text{B}}^{\prime} & L_{\text{B}} \\
m_{\text{B}}      & K_{\text{B}}-m_{\text{B}} & -K_{\text{B}} \\
\end{pmatrix}
\alpha^{l_{\text{B}}l_{\text{B}}^{\prime}}_{m_{\text{B}} K_{\text{B}}-m_{\text{B}}}(0) Q^{l_{\text{A}}}_{m_{\text{A}}}Q^{l_{\text{a}}}_{K_{\text{A}}-m_{\text{A}}},
\end{split}
\end{equation}
\begin{equation}\label{eq:disp}
\begin{split}
C_{n,\text{disp}}^{L_{\text{A}},K_{\text{A}},L_{\text{B}},K_{\text{B}},L}=-
\sum_{l_{\text{A}} l_{\text{A}}^{\prime}l_{\text{B}} l_{\text{B}}^{\prime}}^{{l_{\text{A}}+l_{\text{A}}^{\prime}+l_{\text{B}}+l_{\text{B}}^{\prime}}+2=n}
\zeta^{L_{\text{A}}L_{\text{B}}L}_{l_{\text{A}} l_{\text{A}}^{\prime}l_{\text{B}} l_{\text{B}}^{\prime}}
\sum_{m_{\text{A}}=-l_{\text{A}}}^{l_{\text{A}}}(-1)^{K_{\text{A}}}
\begin{pmatrix}
l_{\text{A}}      & l_{\text{A}}^{\prime} & L_{\text{A}} \\
m_{\text{A}}      & K_{\text{A}}-m_{\text{A}} & -K_{\text{A}} \\
\end{pmatrix}\times \hspace{1.6cm} \\ 
\times \sum_{m_{\text{B}}=-l_{\text{B}}}^{l_{\text{B}}}(-1)^{K_{\text{B}}}
\begin{pmatrix}
l_{\text{B}}      & l_{\text{B}}^{\prime} & L_{\text{B}} \\
m_{\text{B}}      & K_{\text{B}}-m_{\text{B}} & -K_{\text{B}} \\
\end{pmatrix}
8\pi \int_{0}^{\infty}\alpha^{l_{\text{A}}l_{\text{A}}^{\prime}}_{m_{\text{A}} K_{\text{A}}-m_{\text{A}}}(i\omega)
\alpha^{l_{\text{B}}l_{\text{B}}^{\prime}}_{m_{\text{B}} K_{\text{B}}-m_{\text{B}}}(i\omega)
d\omega, 
\end{split}
\end{equation}
\end{widetext}
where the coefficients $\zeta^{L_{\text{A}}L_{\text{B}}L}_{l_{\text{A}} l_{\text{A}}^{\prime}l_{\text{B}} l_{\text{B}}^{\prime}}$ are defined in Ref.~\onlinecite{Rijks:1989_2} and
the formulas are written through the Clebsch-Gordan coupled product of spherical tensors. The irreducible frequency-dependent polarizability $\alpha^{ll^{\prime}}_{mm^{\prime}}(i\omega)$ is defined through the conventional spherical harmonics basis $Y_{lm}$. \cite{Rijks:1989_2} The tensor $\alpha^{ll^{\prime}}_{mm^{\prime}}(i\omega)$ can be transformed to its reducible form $\tilde{\alpha}^{ll^{\prime}}_{mm^{\prime}}(i\omega)$ defined in the basis of the tesseral harmonic cosine and sine functions.  
The induction coefficients $C_{n,\text{ind}}^{L_{\text{A}},K_{\text{A}},L_{\text{B}},K_{\text{B}},L}(\text{A}\rightarrow\text{B})$ define the interaction of the permanent multipole moments of A with the static polarizabilities of B.  The coefficients  $C_{n,\text{ind}}^{L_{\text{A}},K_{\text{A}},L_{\text{B}},K_{\text{B}},L}(\text{B} \rightarrow \text{A})$ are defined in a similar manner  with the interchanged symbols A$\leftrightarrow$B. The total induction coefficients are then defined as the sum of the $\text{B} \rightarrow \text{A}$ and $\text{A} \rightarrow \text{B}$ coefficients. 

The formulas given by \Cref{eq:elst,eq:ind,eq:disp} have been implemented in the \texttt{POLCOR} package. \cite{POLCOR:1992} The calculation of the electrostatic coefficients is rather straightforward, while the dispersion and the induction ones can be calculated using two programs from the package,  \texttt{DISPER} and \texttt{INDUCT}. The programs use the reducible polarizabilities $\tilde{\alpha}^{ll^{\prime}}_{mm^{\prime}}$ and multipole moments $Q^l_m$ as the input. From now on, we skip the over-tilde symbol notation and use $\alpha^{ll^{\prime}}_{mm^{\prime}}$ only for reducible polarizabilities.

When two interacting atoms are in their ground $S$-states, there is no angular dependence $L\!=\!L_{\text{A}}\!=\!L_{\text{B}}\!=\!K_{\text{A}}\!=\!K_{\text{B}}\!=\!0$ and the long-range coefficients are represented only by the dispersion part described by $n$ index, $C_{n,\text{disp}}$ ($n$=6,8,10,$\dots$). More details on this case are given in the supplementary material. 

In the case of a closed-shell nonlinear molecule A and an atom B in the ground $S$-state, such as the complexes of aromatic molecules with alkali-metal and alkaline-earth-metal atoms, the angular dependence is present only for the molecule A, for which we use the notation $L\!=\!L_{\text{A}}\!=\!l$ and $K\!=\!K_{\text{A}}\!=\!m$. We use the notation $C^{l,m}_n$ for such complexes, where $C^{l,m}_n$ is the sum of the dispersion $C^{l,m}_{n,\text{disp}}$ and induction $C^{l,m}_{n,\text{ind}}$ $(\text{A} \rightarrow \text{B})$ coefficients. The $C^{l,m}_{n,\text{ind}}$ $(\text{B} \rightarrow \text{A})$  and $C^{l,m}_{n,\text{elst}}$ vanish for the complexes due to the absence of permanent multipole moments of the atoms. The expression for the long-range dispersion energy, in this case, can be written as follows:
\begin{equation}
E_{\text{disp}}(R,\theta,\phi)=-\sum_{n=6}^{\infty}\sum_{l=0}^{\infty}\sum_{m=-l}^{l}  \frac{C^{l,m}_{n,\text{disp}}}{R^{i}}
\Omega_{lm}(\theta,\phi),
\label{eq:ma}
\end{equation}
where $\Omega_{lm}(\theta,\phi)$ is the tesseral harmonics basis. \cite{Rijks:1989_2}
The maximum values of $n$ and $l$ that are physically relevant are usually taken up to 12 and 6, respectively.  The term with $n\!=\!6$
describes the dipole-dipole  coupling of dynamic polarizabilities of the interacting monomers. The expression for the long-range induction energy is analogous. In this case, the term with $n\!=\!6$ is defined by the coupling of the dipole moment and static dipole-dipole polarizabilities. Additional restrictions on the $l,m$ values stem from the symmetry of  the monomer. For example,  benzene has the $D_{6h}$ point group symmetry, and hence $l\!=0,2,4,\ldots$ and  $m\!=\!0,6,\ldots$ are allowed. Since benzene has  zero dipole moment, the summation  for  $E_{\text{ind}}$
corresponding to that for $E_{\text{disp}}$ in \Cref{eq:ma} starts from $n\!=\!8$. In the case when the monomer has the $C_{2v}$ symmetry like furan, pyridine, and pyrrole, there are no sine-type ($m\!<\!0$) tesseral harmonics in the expansion given by \Cref{eq:ma}.

In order to calculate the sets of the long-range coefficients, one would need to provide the first- and second-order properties for the monomers as the input for the calculation of $C^{l,m}_n$ of the studied complexes employing the approach described above. The theoretical and computational methods employed for the calculation of $\alpha^{ll^{\prime}}_{mm^{\prime}}(i\omega)$ and $Q^l_m$ are explained in the following sections.

\subsection{Alkali-metal and alkaline-earth-metal atoms\label{me}}

The long-range potentials for alkali-metal and alkaline-earth-metal homo- and heterodimers, including their electronically excited states, were reported in numerous previous studies.\cite{Patil:1997,Porsev:2006,Mitroy:2003,Mitroy:2010,Derevianko:2010} However, in Ref.~\onlinecite{Derevianko:2010} only the dipole-dipole polarizabilities $\alpha^{11}(i\omega)$ were tabulated for the given values of $\omega$ on the Gauss-Legendre quadrature grid. Even though the reported results are accurate, \texttt{POLCOR} package employs a different type of quadrature -- the Chebyshev-Gauss one.  Moreover, one would also need $\alpha^{ll}(i\omega)$ for $l\!>\!1$, which were never provided in numerical form in the literature. Therefore, we decided to calculate the polarizabilities $\alpha^{ll}(i\omega)$ with $l$ up to 4 for the metal atoms. 
 
\begin{table}[tb!]
\caption{The values of the static electric dipole and quadrupole polarizabilities $\alpha^{ll}$ ($l=1,2$) for alkali-metal and alkaline-earth-metal atoms and the long-range dispersion coefficients $C_{n}=C_{n,\text{disp}}$ ($n=6,8$) for their homodimers. The upper rows contain our results, and the lower ones the results from Ref.~\onlinecite{Derevianko:2010} for $\alpha^{11}$ and $C_{6}$ and from Ref.~\onlinecite{Mitroy:2003} for $\alpha^{22}$ and $C_{8}$ (except for Cs and Ba). All values are in atomic units.}
\label{tab:me}
\begin{ruledtabular}
\begin{tabular}{ccccc}
Atom/Dimer & $\alpha^{11}(\times 10^2)$  & $\alpha^{22}(\times 10^3)$  & $C_{6}(\times 10^3)$ & $C_{8}(\times 10^4)$\\
\hline 
Li    &1.648&1.393&1.395 &8.148\\
      &1.640& 1.424&1.389&8.352 \\
Na    &1.608&1.796&1.506& 10.847\\
      &1.626&1.881&1.556&  11.600\\
K     &2.897& 4.703&3.750&38.159\\
      &2.902&5.018&3.897&  42.070\\
Rb    &3.195&6.068&4.456& 51.199\\
      &3.186&6.480&4.690&  52.701\\
Cs    &4.041&10.259&6.434& 91.825\\
      &3.998&\enspace 10.470\footnote{Results from Ref.~\onlinecite{Porsev:2003}.\label{note:Porsev:2003}}&6.846 & \enspace 102\textsuperscript{\ref{note:Porsev:2003}}\\   
Be    &0.377&0.297&0.214&1.013\\
      &0.378&0.301&0.214&1.022 \\
Mg    &0.717&0.809&0.638&4.171\\
      &0.713&0.814&0.627&4.164 \\
Ca    &1.574&3.035&2.193& 22.500\\
      &1.571&3.063&2.121& 22.60\\
Sr    &1.962&4.477&3.207&37.753\\
      &1.972&4.577& 3.103& 38.54\\
Ba    &2.710&8.225&5.413& 77.28\\
      &2.735&\enspace8.900\footnote{Results from Ref.~\onlinecite{Porsev:2006}.\label{note:Porsev:2006}}&5.160 &77.2\textsuperscript{\ref{note:Porsev:2006}}\\                                 
\end{tabular}
\end{ruledtabular}
\end{table} 
 
First, we have implemented the theoretical approach reported in Ref.~\onlinecite{Patil:1997} for the calculation of the static and dynamic multipole polarizabilities of alkali-metal atoms.  Within this approach, the approximate analytical wavefunctions for the systems with one valence electron are used. Such an approach is convenient because it requires no additional electronic structure calculations and gives good agreement with other theoretical methods.  The \texttt{Maple} software package was used for that purpose.  

The alkaline-earth-metal atoms are closed-shell, which simplifies the calculation of their properties using the implemented electronic structure methods. We used the explicitly connected representation of the expectation value of a one-electron operation and CCSD polarization propagator.\cite{Moszynski:2005, Korona:2006_2} For heavy Sr and Ba, this method was used together with the relativistic spin-free Douglas-Kroll-Hess (DKH) Hamiltonian. \cite{Hess:1986, Jansen:1989} The core-electron correlation effects were taken into account in these calculations. It was shown previously in Refs. \onlinecite{Misquitta:2005_2,Hesselmann:2013, Shirkov:2015_1,Shirkov:2017} for noble-gas atoms that the second-order properties are sensitive to the presence of additional diffuse functions in the basis sets. Therefore,  in order to obtain the values of $\alpha^{ll}(i\omega)$ close to the convergence limit, we used the triply augmented Dunning's basis sets with diffuse even-tempered functions and additional tight core-correlated functions -- t-aug-cc-pwCV5Z  for Be and Mg,\cite{Prascher:2010} t-aug-cc-pwCVQZ for Ca, t-aug-cc-pwCVQZ-X2C for Sr and t-aug-cc-pwCVTZ-X2C for Ba.\cite{Hill:2017}  One has to note that the t-aug-cc-pwCVQZ-X2C basis set is available for Ba, but we experienced some convergence problems with this basis set and used the smaller t-aug-cc-pwCVTZ-X2C one.  

\Cref{tab:me} presents the calculated values of the static electric dipole and quadrupole polarizabilities $\alpha^{ll}$, $l=1,2$,  for alkali-metal and alkaline-earth-metal atoms as well as the dispersion coefficients $C_{n,\text{disp}}$ with $n=6,8$ for their homodimers.  The $C_{n,\text{disp}}$ coefficients were calculated at an excessive quadrature grid with $N\!=\!50$ points to guarantee the convergence of the numerical integration.  As one can see in \Cref{tab:me}, the agreement is better than 3\% for most of the quantities confirming our good choice of the methods used.  One has to note that our results for alkali atoms are slightly different from those obtained in Ref.~\onlinecite{Patil:1997}, because we did not use any approximations to the hypergeometrical function.  The calculated values of $\alpha^{ll}(i\omega)$ for $l=1,2,3,4$ at the grid of $i\omega$ as well as the higher-order $C_{n,\text{disp}}$ coefficients for the homodimers with $n$ up to 12 are given in the supplementary material.

\begin{table}[tb!]
\caption{The values of the electric dipole moment $\mu$, the components of the electric quadrupole moment $Q^2_m$ and static electric dipole-dipole polarizabilities $\alpha^{11}_{mm'}$ for the studied aromatic molecules. The first row contains DFT response results, and the second one the finite-field CCSD(T) results. All values are in atomic units.} 
\label{tab:arom}
\begin{ruledtabular}
\begin{tabular}{ccccccc}
Molecule & $\mu$ & $Q^2_0$ & $Q^2_2$  &$\alpha^{11}_{00}$   &$\alpha^{11}_{11}$ &$\alpha^{11}_{\textrm{-}1\textrm{-}1}$\\
\hline
benzene  &-&-5.83&-&44.41& 81.19&81.19 \\
         &-&-5.82&-&44.29&79.20&79.20 \\
pyridine &0.93&-3.74&-4.54&40.61&77.84&73.24 \\
         &0.89&-3.73&-4.53&40.41&75.87&71.64  \\
furan    &-0.25&-4.38&-2.41&34.69&58.80&53.01\\
         &-0.26&-4.35& -2.51&34.48&56.65&52.70  \\
pyrrole  &0.73&-6.36&2.09&37.04&60.94&64.26 \\
         &0.72&-6.38&2.03&38.55&60.54&62.67 \\
\end{tabular}
\end{ruledtabular}
\end{table} 

\begin{table*}[t]
\caption{The selected values of the long-range dispersion $C^{l,m}_{n,\text{disp}}$ and induction $C^{l,m}_{n,\text{ind}}$ coefficients (in atomic units) for the complexes of aromatic molecules with alkali-metal atoms.}
\label{tab:res_am}
\begin{ruledtabular}
\begin{tabular}{ccccccc}
Complex& $C^{0,0}_{6,\text{disp}}(\times 10^3)$  & $C^{0,0}_{6,\text{ind}}(\times 10^2)$& $C^{2,0}_{6,\text{disp}}(\times 10^2)$ &  $C^{2,0}_{6,\text{ind}}(\times 10^2)$ & $C^{0,0}_{8,\text{disp}}(\times 10^5)$& $C^{0,0}_{8,\text{ind}}(\times 10^4)$\\
\hline
benzene-&&&&&&\\
Li  &1.018&-&-3.853&-&0.980&0.839\\
Na  &1.103&-&-4.156&- &1.116&0.819\\
K   &1.583&-&-6.026& -&1.875&1.475\\
Rb  &1.710&-&-6.516& -&2.117&1.627\\ 
Cs  &1.979&-&-7.564& -&2.661&2.058\\
\hline
pyridine-&&&&&&\\
Li  &0.946&1.426&-3.671& -1.594&0.839&1.157\\
Na  &1.026&1.391&-3.961& -1.555&1.019&1.223\\
K   &1.472&2.505&-5.739& -2.801&1.715&2.521\\
Rb  &1.590&2.762&-6.206& -3.089&1.937&2.971\\ 
Cs  &1.840&3.494&-7.202& -3.907&2.438&4.316\\   
\hline
furan-&&&&&&\\
Li  &0.730&0.091&-2.278& -0.111&0.626&0.638\\
Na  &0.791&0.097&-2.463& -0.108&0.716&0.629\\
K   &1.133&0.174&-3.552& -0.195&1.222&1.156\\
Rb  &1.224&0.192&-3.840&-0.215&1.385&1.288\\ 
Cs  &1.415&0.241&-4.451&-0.271&1.753&1.668\\   
\hline
pyrrole-&&&&&&\\
Li  &0.811&0.871&-2.555&-0.182&0.706&1.138\\
Na  &0.879&0.850&-2.761& -0.950&0.808&1.318\\
K   &1.260&1.530&-3.987& -1.711&1.376&2.567\\
Rb  &1.362&1.688&-4.310& -1.887&1.560&2.948\\ 
Cs  &1.575&2.134&-4.996& -2.386&1.972&4.069\\          
\end{tabular}
\end{ruledtabular}
\end{table*} 

\subsection{Aromatic molecules\label{arom}}

As the next step, we calculated the first- and second-order electric properties of the selected aromatic molecules using their available empirical and theoretical equilibrium geometries.\cite{Nygaard:1969,Sorensen:1974,Larsen:1979,Gauss:2000}  The components of the multiple moments and polarizabilities may depend on the choice of the coordinate frame.  We put the center of frame to the monomers center of mass, putting the N or O atoms (C for benzene) on the $x$-axis. The $x$-coordinates of N atoms in pyridine and pyrrole are positive, and the coordinate of O atom in furan is negative. The values of the Cartesian geometries of the molecules are given in the supplementary material. 

The electronic structure calculations for such large molecules are challenging and rather time-consuming. Therefore, our choice of method must be a compromise between accuracy and computation cost. The static properties can be calculated using the finite field (FF) approach.\cite{Cohen:1965} In such a way, the values of the electric dipole $\mu=Q^1_1$, non-zero components of the electric quadrupole moment, and electric dipole-dipole static polarizabilities were first calculated using FF with the CCSD(T) method. The augmented Dunning's quadruple zeta basis set (aug-cc-pVQZ)~\cite{DunningJCP89} and the value of the external field perturbation equal to 0.001 a.u. were used for these calculations. The calculated properties are collected in \Cref{tab:arom} and can be considered as the benchmark.  However, such an approach becomes inconvenient for higher orders of multipole moments and polarizabilities due to the dependence on the value of the field increment and the accuracy of the finite difference formulas used. Moreover, one also needs the values of dynamic polarizabilities, which cannot be calculated by this method.

\begin{table*}[t]
\caption{The selected values of the long-range dispersion $C^{l,m}_{n,\text{disp}}$ and induction $C^{l,m}_{n,\text{ind}}$ coefficients (in atomic units) for the complexes of aromatic molecules with alkaline-earth-metal atoms.}
\label{tab:res_aem}
\begin{ruledtabular}
\begin{tabular}{ccccccc}
Complex& $C^{0,0}_{6,\text{disp}}(\times 10^3)$  & $C^{0,0}_{6,\text{ind}}(\times 10^2)$& $C^{2,0}_{6,\text{disp}}(\times 10^2)$ &  $C^{2,0}_{6,\text{ind}}(\times 10^2)$ & $C^{0,0}_{8,\text{disp}}(\times 10^5)$& $C^{0,0}_{8,\text{ind}}(\times 10^4)$\\
\hline
benzene-&&&&&&\\
Be  &0.563&-&-2.006& -&0.454&0.192\\
Mg  &0.932&-&-3.360& -&0.838&0.365\\
Ca  &1.580&-&-5.725& -&1.731&0.801\\
Sr  &1.899&-&-6.875& -&2.199&0.999\\
Ba  &2.416&-&-8.758& -&3.038&1.380\\ 
\hline
pyridine-&&&&&&\\
Be  &0.512&0.327&-1.859&-0.365&0.414&0.260\\
Mg  &0.851&0.620&-3.117& -0.694&0.764&0.547\\
Ca  &1.474&1.361&-5.465& -1.521&1.583&1.473\\
Sr  &1.771&1.696&-6.563& -1.897&2.012&1.986\\
Ba  &2.254&2.344&-8.359& -2.621&2.784&3.186\\   
\hline
furan-&&&&&&\\
Be  &0.400&0.023&-1.184& -0.025&0.285&0.146\\
Mg  &0.660&0.043&-1.976&-0.048&0.532&0.281\\
Ca  &1.142&0.094&-3.443&-0.106&1.123&0.635\\
Sr  &1.373&0.118&-4.134& -0.132&1.434&0.802\\
Ba  &1.745&0.163&-5.261& -0.182&1.996&1.139\\ 
\hline
pyrrole-&&&&&&\\
Be  &0.441&0.200&-1.322& -0.223&0.321&0.293\\
Mg  &0.730&0.379&-2.207& -0.424&0.599&0.589\\
Ca  &1.263&0.831&-3.851& -0.930&1.261&1.458\\
Sr  &1.519&1.036&-4.624& -1.158&1.609&1.909\\
Ba  &1.932&1.432&-5.885&-1.600&2.240&2.907\\       
\end{tabular}
\end{ruledtabular}
\end{table*}

Therefore, we used the density functional theory (DFT) response approach for the calculation of the static and dynamic properties, $Q^l_m$ and $\alpha^{ll'}_{mm'}(i\omega)$ with $l$ up to 4. \cite{Osinga:1997,Hesselmann:2013,Hesselmann:2015} The density fitting approximation with the LPBE0ac exchange-correlation functional and the adiabatic local-density approximation (ALDA)\cite{Hesselmann:2003_1,Misquitta:2003,Hesselmann:2005,Podeszwa:2006_2} kernel together with the gradient regulated asymptotic correction (AC) with the Leeuwen-Baerends exchange-correlation potential was employed for the DFT response calculations. \cite{Tozer:1998,Gruning:2001} We previously used this functional for the calculation of the interaction energies for the series of complexes of  aromatic molecules with Ar atoms using DFT-SAPT. \cite{Makarewicz:2016,Shirkov:2019} The following values of the AC were taken: 0.072, 0.075, 0.072, and 0.077 a.u.~for benzene, pyridine, furan, and pyrrole, correspondingly.
We calculated these corrections  as the difference of negative LPBE0 HOMO energies and
the PBE0 ionization potentials.  The atomic orbital basis sets aug-cc-pVQZ with auxiliary aug-cc-pVQZ-JKFIT were employed for these calculations.\cite{Weigend:2002_1,Weigend:2002_2} Unfortunately, we were not able to use the larger basis set for these calculations due to convergence problems and linear dependence of the basis set functions. Therefore, one can expect that the values of $\alpha^{ll'}_{mm'}(i\omega)$ may be not fully converged, especially for higher $l$ values. The grid of  $(i\omega)$ was taken the same as in \Cref{me} for alkali and alkaline-earth-metal atoms.  

The comparison of the static properties calculated with the DFT response and the CCSD(T) benchmark is given in \Cref{tab:arom}.  As one can see, the deviation of the DFT response results from their CCSD(T) counterparts does not exceed 2-3\% for most of the values being somewhat higher for pyridine.  The full sets of calculated values of $Q^l_m$  and  $\alpha^{ll'}_{mm'}(i\omega)$ with $l$ up to 4 are given in the supplementary material.  

The developer's version of the \texttt{Molpro} package was used for the calculations of the properties of the alkaline-earth-metal atoms and the aromatic molecules.\cite{MOLPRO}

\section{Results\label{res}}

The selected values of the leading long-range dispersion $C^{l,m}_{n,\text{disp}}$ and induction $C^{l,m}_{n,\text{ind}}$ coefficients are presented in \Cref{tab:res_am,tab:res_aem} for the complexes of aromatic molecules with alkali-metal and alkaline-earth-metal atoms, respectively. The remaining coefficients with $n$ up to 12 and $l$ up to 6 are given in the supplementary material. We have to note that the coefficients with $n\!=\!11,12$ and $m\!\neq\!0$ also depend on the first- and second-order properties with $l\!>\!4$, which were not used in this work. However, it is expected that the contribution from these neglected higher quantities is minor.  

Our values of $C^{6,0}_{\text{disp}}$ and $C^{6,2}_{\text{disp}}$ for the complexes with benzene are very close to those obtained in Ref.~\onlinecite{Wojcik:2019}. However, our values $C^{6,2}_{\text{disp}}$ are two times larger due to a different normalization factor used for the spherical harmonics.  The complexes with aromatic molecules with a non-zero dipole moment acquire the coefficients $C^{l,m}_{n}$ with odd $n$ values. Thus, $C^{7,1}_{\text{disp}}$ describes the interactions of the anisotropic part $l\!\neq\! l'$ of the dipole-dipole polarizability  $\alpha^{ll'}_{mm'}$ of an aromatic molecules with $\alpha^{11}$ of a metal atom.  

\begin{figure*}[t]  
\begin{center}
\caption{{\footnotesize Plots of  the functions  $f_{n_{\text{m}}}(R)=E^{n_{\text{m}}}_{\text{disp}}(R,\theta,\phi)/E^{12}_{\text{disp}}(R,\theta,\phi)$ ($n_{\text{m}}=6,8,10$) for the pyridine-Li complex at different geometries given by fixed values of
 $(\theta,\phi)$ (in degrees).}}
\label{fig:pyrli}
\includegraphics[angle=-90,width=8.3cm]{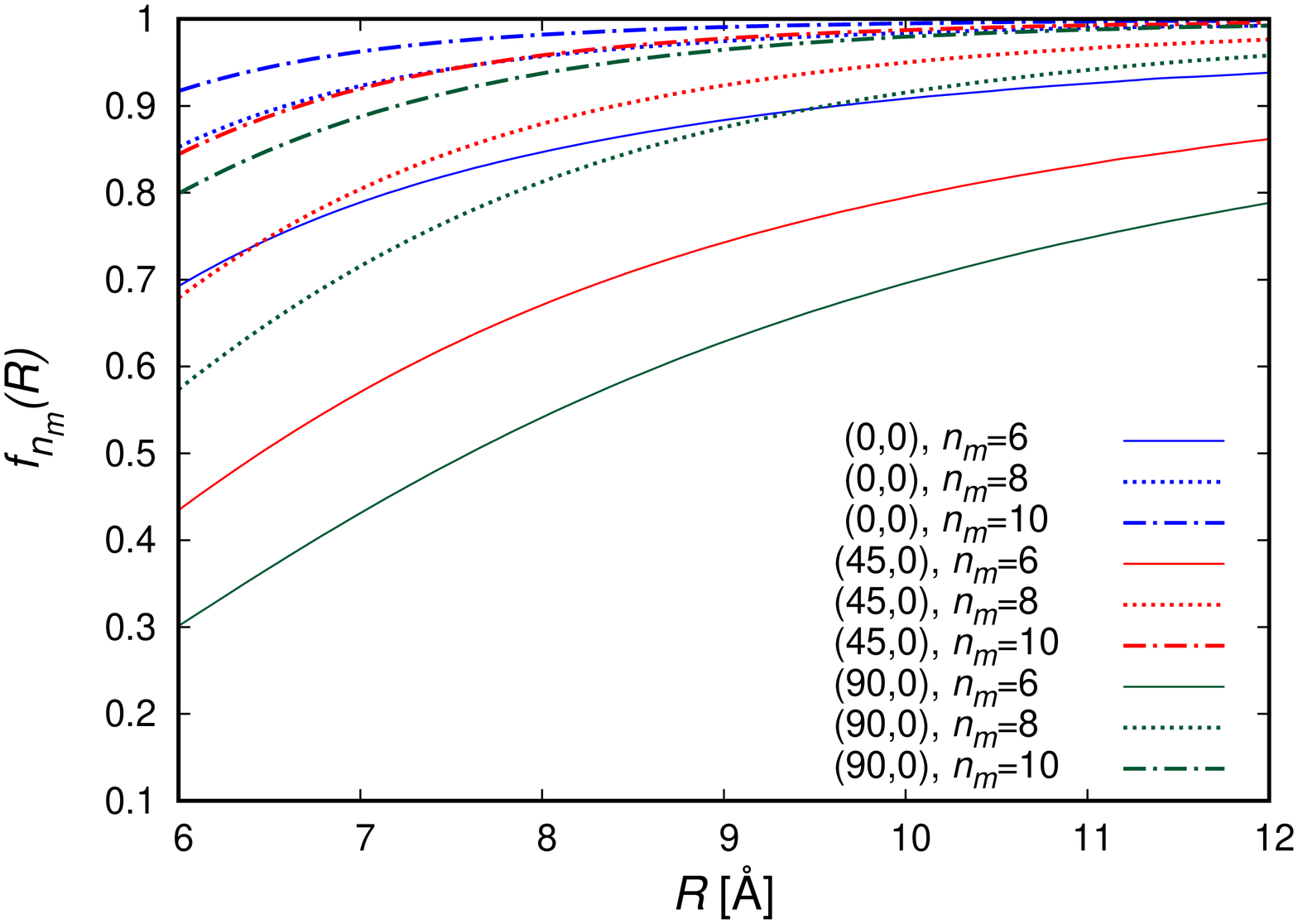} \includegraphics[angle=-90,width=8.3cm]{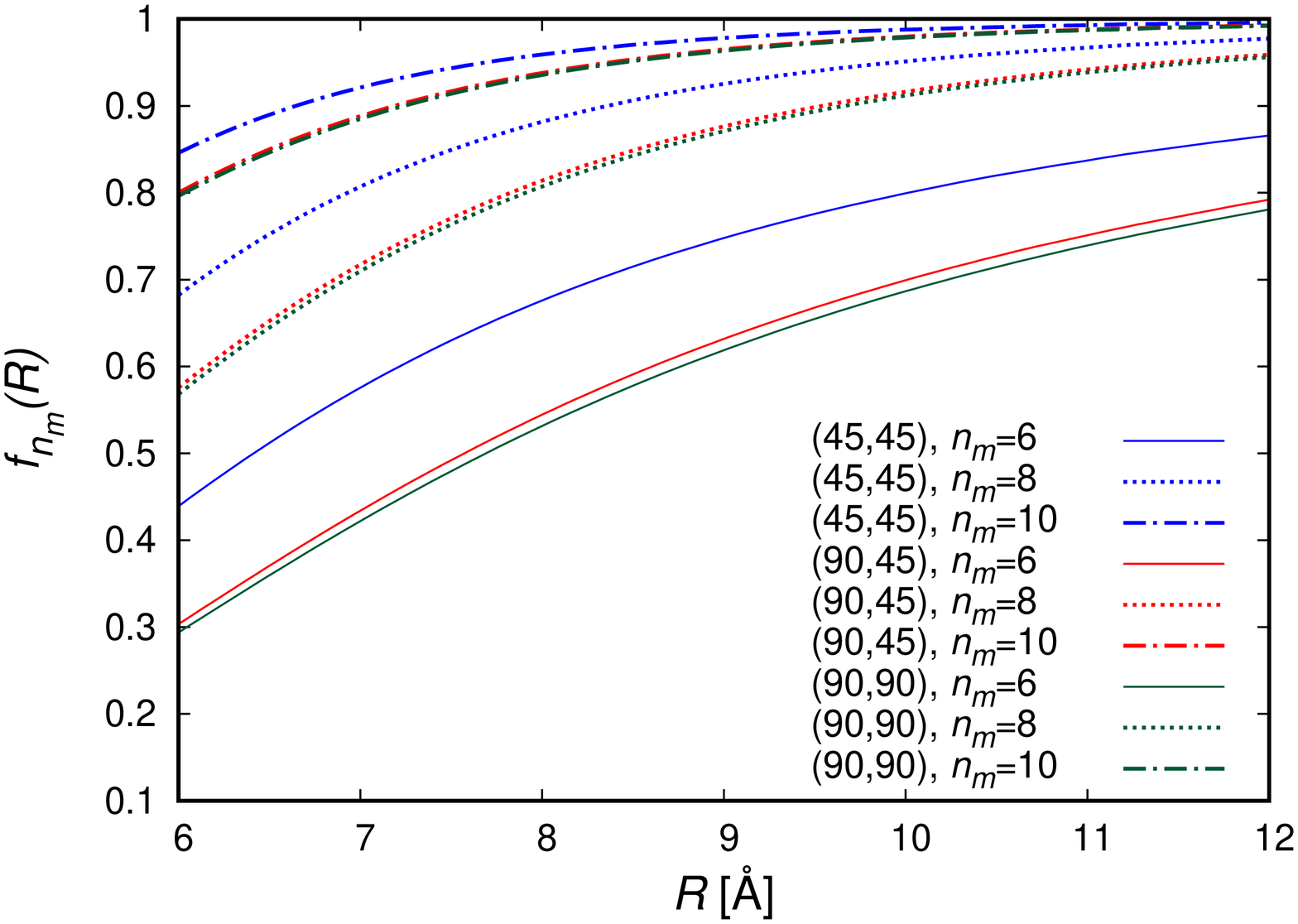}  \\
\end{center}
\end{figure*}

\begin{figure*}[t]  
\begin{center}
\caption{{\footnotesize Plots of  the functions  $f_{n_{\text{m}}}(R)=E^{n_{\text{m}}}_{\text{disp}}(R,\theta,\phi)/E^{12}_{\text{disp}}(R,\theta,\phi)$ ($n_{\text{m}}=6,8,10$) for the furan-Sr complex at different geometries given by fixed values of
 $(\theta,\phi)$ (in degrees).}}
\label{fig:fursr}
\includegraphics[angle=-90,width=8.3cm]{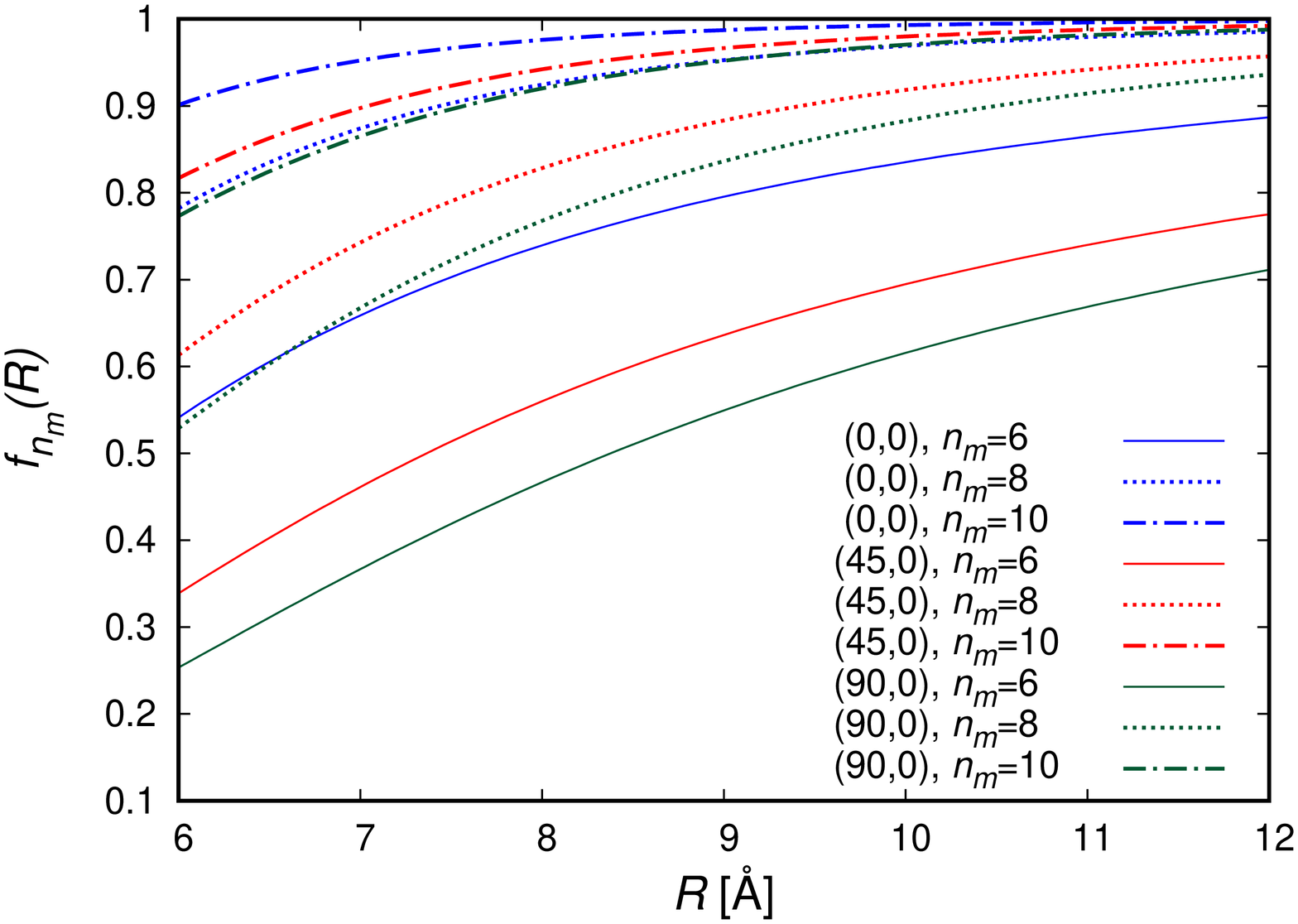} \includegraphics[angle=-90,width=8.3cm]{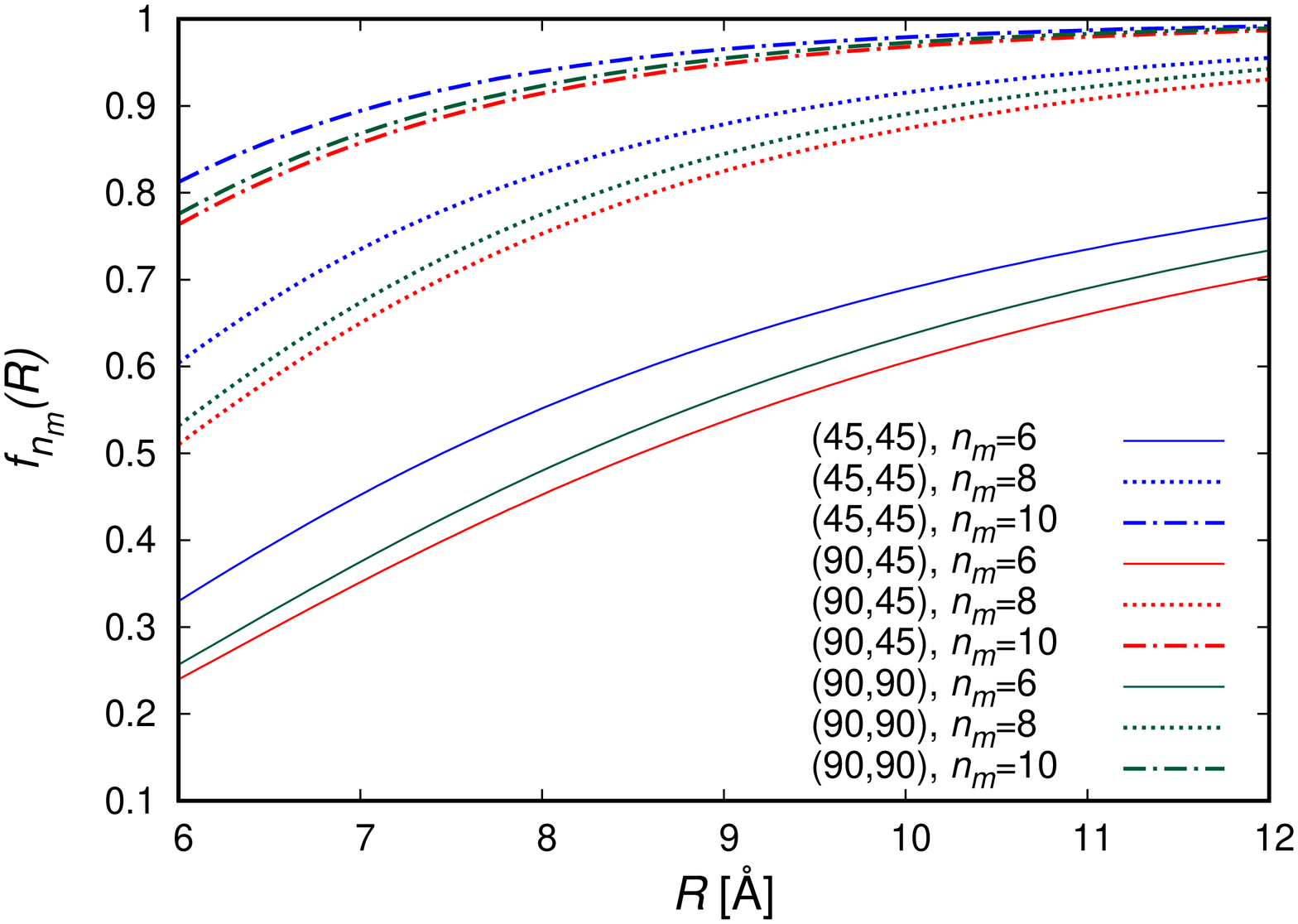}  \\
\end{center}
\end{figure*}

The absolute values of the coefficients presented in \Cref{tab:res_am,tab:res_aem} grow as the atomic number of the interacting metal atoms grows in each group of elements, with a few exceptions for Li and Na. This can be explained by the fact that the dipole-dipole polarizability of Na is slightly lower than that of Li, see \Cref{tab:me}. A comparison of the induction coefficients $C^{l,m}_{n,\text{ind}}$ reveals that the coefficients for the complexes with alkali-metal atoms are larger than for the complexes with alkaline-earth-metal atoms from the same row of elements. This is in agreement with the fact that the alkali-metal atoms have higher polarizabilities than the corresponding alkaline-earth-metal atoms. However, there is no such regularity for the dispersion coefficients $C^{l,m}_{n,\text{disp}}$, because these coefficients also depend on the behavior of the frequency-dependent functions $\alpha^{ll'}_{mm'}(i\omega)$.

\Cref{fig:pyrli,fig:fursr} presents the results for the pyridine-Li complex and furan-Sr complexes, respectively, where the comparison
of the long-range dispersion energy $E_{\text{disp}}^{n_{\text{m}}}(R,\theta,\phi)$ found with different values of $n_{\text{m}}$ -- the maximum
power of the terms $1/R^{n}$ in \Cref{eq:ma} -- are plotted at different values of angles $(\theta,\phi)$. As one can see, the terms with only  $n\!=\!6$ are up to three times smaller than their counterparts with $n_{\text{m}}\!=\!12$ in the region $R\sim 6$ \AA.{ }The largest deviation is observed at $\theta\!=\!90^{\circ}$. The difference between $n_{\text{m}}\!=\!10$ is $n_{\text{m}}\!=\!12$ is hardly noticeable even for smaller $R$ values.  

A \texttt{Fortran 90} routine for the calculation of the long-range interaction energy $E_{\text{int}}=E_{\text{disp}}^{(2)}+E_{\text{ind}}^{(2)}$ from the the sets of $C^{l,m}_{n}=C^{l,m}_{n,\text{disp}}+C^{l,m}_{n,\text{ind}}$  is available  in the supplementary material.

\section{Conclusions and future prospects\label{conc}}

We have presented the long-range potentials for a series of complexes containing an aromatic molecule (benzene, pyridine, furan, and pyrrole) with alkali-metal and alkaline-earth-metal atoms. In order to obtain these potentials, we not only reproduced the second-order properties of alkali-metal atoms using an analytical approach but also calculated these properties for alkaline-earth-metal atoms using an accurate {\slshape ab initio} method based on the expectation-value coupled cluster theory. We have also shown that the DFT response method reproduces the static first- and second-order  properties of the aromatic molecules with high accuracy. This has been done by the comparison of the DFT response results with their CCSD(T) counterparts calculated using the finite field approach for the static properties. 

The sets of the long-range dispersion $C^{l,m}_{n,\text{disp}}$ and induction $C^{l,m}_{n,\text{ind}}$ coefficients have been calculated using the available first principles analytical formulas. It has been shown that it is important to include the terms with $n\!>\!6$ for a more accurate description of the interaction energy in the van der Waals region. The reported long-range potentials can complement the potentials used to study the dynamics of the considered complexes. Further study will include calculating the vibrational bound states and cross sections of elastic and inelastic scattering using the implemented theoretical approaches. \cite{Shirkov:2015_2,Shirkov:2020,MOLSCAT:2020}

The proposed approach applies not only to other numerous complexes of single-ring aromatic molecules with alkali-metal and alkaline-earth-metal atoms but also to other polyatomic molecules prospective for ultracold cooling. An interesting extension of the approach applicable for longer polycyclic molecules such as tetracene and anthracene might be multi-center expansion. \cite{Moszynski:2007} However, such an approach would require some sophisticated angular momentum algebra. The complexes with large aromatic molecules also revealed another type of problem -- the need for multireference wavefunctions for the calculation of the supermolecular interaction energy in some regions of the PES.\cite{Wojcik:2019}. 
   
Finally, an accurate description of the long-range intermolecular interactions is very important for ultracold physics and chemistry studies.\cite{QuemenerCR12} Scattering properties at ultralow temperatures are very sensitive to tiny details of the PES,\cite{GronowskiPRA20} including interactions at large distances, where quantum tunneling and reflection from a centrifugal barrier happen. Therefore, we expect our results to be useful in developing and investigating theoretically sympathetic cooling and controlled chemical reactions of polyatomic functionalized aromatic molecules\cite{AugenbraunJPCL22,MitraJPCL22} immersed in ultracold atomic gases, which will be realized in upcoming experiments aiming at the formation and application of polyatomic molecules at ultralow temperetures.

\section*{Supplementary Material}
The supplementary material contains the following data~\cite{supplemental}:
\begin{itemize}[noitemsep]
  \item A \texttt{Maple} routine for generating the static and dynamic polarizabilities of alkali-metal atoms for any value of $l$ and $\omega$. 
  \item Calculated static and dynamic multipole polarizabilities for $l$ up to 4 at the given grid points of $\omega$ for alkali-metal and alkaline-earth-metal atoms. 
  \item Values of multipole moments and static and dynamical polarizabilities for the studied aromatic molecules for $l$ up to 4. 
  \item Dispersion coefficients $C_{n,\text{disp}}$ for alkali-metal and alkaline-earth-metal homodimers with $n$ up to 12. 
  \item Sets of $C^{l,m}_{n,\text{disp}}$ and $C^{l,m}_{n,\text{ind}}$ for the complexes of  aromatic molecule with metal atoms with $n$ up to 12. 
  \item A \texttt{Fortran 90} routine for converting the calculated sets of long-range $C^{l,m}_{n}$ coefficients to the interaction energy $E_{\text{int}}$.    
\end{itemize}

\section*{Acknowledgments}
Financial support from the National Science Center Poland (Grant No.~2015/19/D/ST4/02173) is gratefully acknowledged. The computational part of this research has been partially supported by the PL-Grid Infrastructure.

\bibliography{biblio}
  
\end{document}